\RequirePackage[2020-02-02]{latexrelease}
\documentclass
[aps,twocolumn,prl,superscriptaddress,amsmath,tightenlines,twoside,bibnotes]{revtex4}%

\usepackage{amsfonts}
\usepackage{amsmath}
\usepackage{amssymb}
\usepackage{version}
\usepackage[truetex]{graphicx}
\usepackage[truetex]{color}
\usepackage{xcolor}
\usepackage{hyperref}
\usepackage{float}%

\begin{document}

\title{Induced transparency: interference or polarization?}
\author{Changqing Wang}
\affiliation{Department of Electrical and Systems Engineering, Washington University, St Louis, MO, 63130, USA.}
\author{Xuefeng Jiang}
\affiliation{Department of Electrical and Systems Engineering, Washington University, St Louis, MO, 63130, USA.}
\author{William R. Sweeney}%
\affiliation{Departments of Applied Physics and Physics, Yale University, New Haven, CT, 06520, USA.}
\affiliation{Yale Quantum Institute, Yale University, New Haven, CT 06520, USA.}
\author{Chia Wei Hsu}
\affiliation{Ming Hsieh Department of Electrical Engineering, University of Southern California, Los Angeles, California 90089, USA}
\author{Yiming Liu}
\affiliation{Department of Electrical and Systems Engineering, Washington University, St Louis, MO, 63130, USA.}
\author{Guangming Zhao}
\affiliation{Department of Electrical and Systems Engineering, Washington University, St Louis, MO, 63130, USA.}
\author{Bo Peng}
\altaffiliation[Present address: ]{Globalfoundries, East Fishkill, NY, USA.}
\affiliation{Department of Electrical and Systems Engineering, Washington University, St Louis, MO, 63130, USA.}
\author{Mengzhen Zhang}%
\affiliation{Departments of Applied Physics and Physics, Yale University, New Haven, CT, 06520, USA.}
\affiliation{Yale Quantum Institute, Yale University, New Haven, CT 06520, USA.}
\author{Liang Jiang}%
\affiliation{Pritzker School of Molecular Engineering, University of Chicago, Chicago, IL 60637, USA.}
\author{A. Douglas Stone}%
\affiliation{Departments of Applied Physics and Physics, Yale University, New Haven, CT, 06520, USA.}
\affiliation{Yale Quantum Institute, Yale University, New Haven, CT 06520, USA.}
\author{Lan Yang}%
\email[Corresponding author. ]{Email: yang@seas.wustl.edu}
\affiliation{Department of Electrical and Systems Engineering, Washington University, St Louis, MO, 63130, USA.}

\keywords{electromagnetically induced transparency, interference, polarization}
\begin{abstract}
The polarization of optical fields is a crucial degree of freedom in the all-optical analogue of electromagnetically induced transparency (EIT). However, the physical origins of EIT and polarization induced phenomena have not been well distinguished, which can lead to confusion in associated applications such as slow light and optical/quantum storage.
Here we study the polarization effects in various optical EIT systems.
We find that a polarization mismatch between whispering gallery modes in two indirectly coupled resonators can induce a narrow transparency window in the transmission spectrum resembling the EIT lineshape.
However, such polarization induced transparency (PIT) is distinct from EIT: it originates from strong polarization rotation effects and shows unidirectional feature.
The coexistence of PIT and EIT provides new routes for the manipulation of light flow in optical resonator systems.

\end{abstract}

\maketitle

Coherent processes of light-matter interaction have been known to generate
electromagnetically induced transparency (EIT) in optical media with
$\Lambda$-shape energy levels \cite{fleischhauer2005electromagnetically, marangos1998electromagnetically}. The probability of occupation on the excited state is cancelled due to the destructive interference between two excitation pathways and thus the
absorption of a probe beam is annihilated. Associated with EIT is the
strong normal dispersion and group delay, which play a critical role in
applications of slow light and optical storage \cite{lukin2001controlling,wu2010slow,hau1999light,zhou2013slowing,beck2016large}. EIT has been widely studied in
atomic systems \cite{rohlsberger2012electromagnetically,harris1990nonlinear,boller1991observation}, superconductors \cite{anisimov2011objectively,abdumalikov2010electromagnetically}, electronics \cite{garrido2002classical}, metamaterial/metasurfaces \cite{papasimakis2008metamaterial,tassin2012electromagnetically}, optical resonators \cite{xu2007breaking,yanik2004stopping,limonov2017fano,totsuka2007slow,liu2017electromagnetically,dong2015brillouin,kim2015non}, scattering nanostructures \cite{hsu2014theoretical},
optomechanics \cite{safavi2011electromagnetically,weis2010optomechanically,dong2012optomechanical,lu2018optomechanically}, plasmonics \cite{liu2009plasmonic,dyer2013induced,taubert2012classical}, etc. Among them, coupled-mode optical platforms, including a single resonator \cite{li2011experimental,zhao2017raman,zheng2016optically,xiao2009electromagnetically}, directly coupled microresonators \cite{peng2014and} and
indirectly coupled microresonators \cite{li2012experimental,xu2006experimental}, have been
intensively explored as a promising candidate for realizing the all-optical analogue of EIT, due to the merit of room temperature operation, on-chip
integratability, and high tunability for parameter control. The recent study of exceptional-point-assisted transparency (EPAT) \cite{EITEP2019Wang} offers opportunities for EIT control via chiral eigenstates associated with the exceptional points (EPs) \cite{EITEP2019Wang,miri2019exceptional,wang2021non,chen2021non,wiersig2014chiral,peng2016chiral,miao2016orbital,sweeney2019perfectly,wang2021coherent}. 
While there have been comprehensive explorations into the roles of intermodal
coupling, resonance frequencies, optical dissipation rates and phase factors of
propagation, the investigation of another important degree of
freedom --- polarization states of probe fields and optical modes --- have been lacking. It
is known that the probe and coupling field polarizations have significant influence on the magnitude of EIT in
multilevel cascade atomic systems \cite{wielandy1998coherent,drampyan2009electromagnetically,wang2006controlling,mcgloin2000polarization}. In optical systems, the polarization mismatch between the mode fields in different optical devices naturally exists, and the transmission lineshape was found to be modified by the polarization of incident light \cite{li2011experimental}. Moreover, transparency and absorption phenomena can occur in a single resonator supporting overlapping modes with different polarizations \cite{rosenberger2013eit,rosenberger2016comparison,bui2016experimental}. However, up to now, the distinction between EIT and polarization effects is unclear in two ways: 1) how is EIT affected by the polarization mismatch in different coupled-mode optical systems, and 2) what is the underlying physics of the transparency phenomena caused by polarization effects. To be able to clearly understand the polarization effects and their distinction from EIT is of great significance for properly controlling and utilizing polarization in the induced transparency phenomena for applications in slow light generation, optical switching, sensing, etc.

Here we report a thorough study on the effects of polarization in
various configurations for the all-optical analogue of EIT. In particular, we find the polarization induced transparency (PIT) phenomenon in indirectly coupled resonators, which exhibits a unidirectional feature. This phenomenon is found to be strongly dependent on the polarization mismatch between two cavity modes.
Moreover, by exploiting backscattering on the resonator surfaces, the indirectly coupled resonators can function as a hybrid system that involves EIT and PIT simultaneously.

\section*{Polarization effects on EIT in various configurations}

\begin{figure*}%[tptb]
\centering
\includegraphics[width=11.4cm]{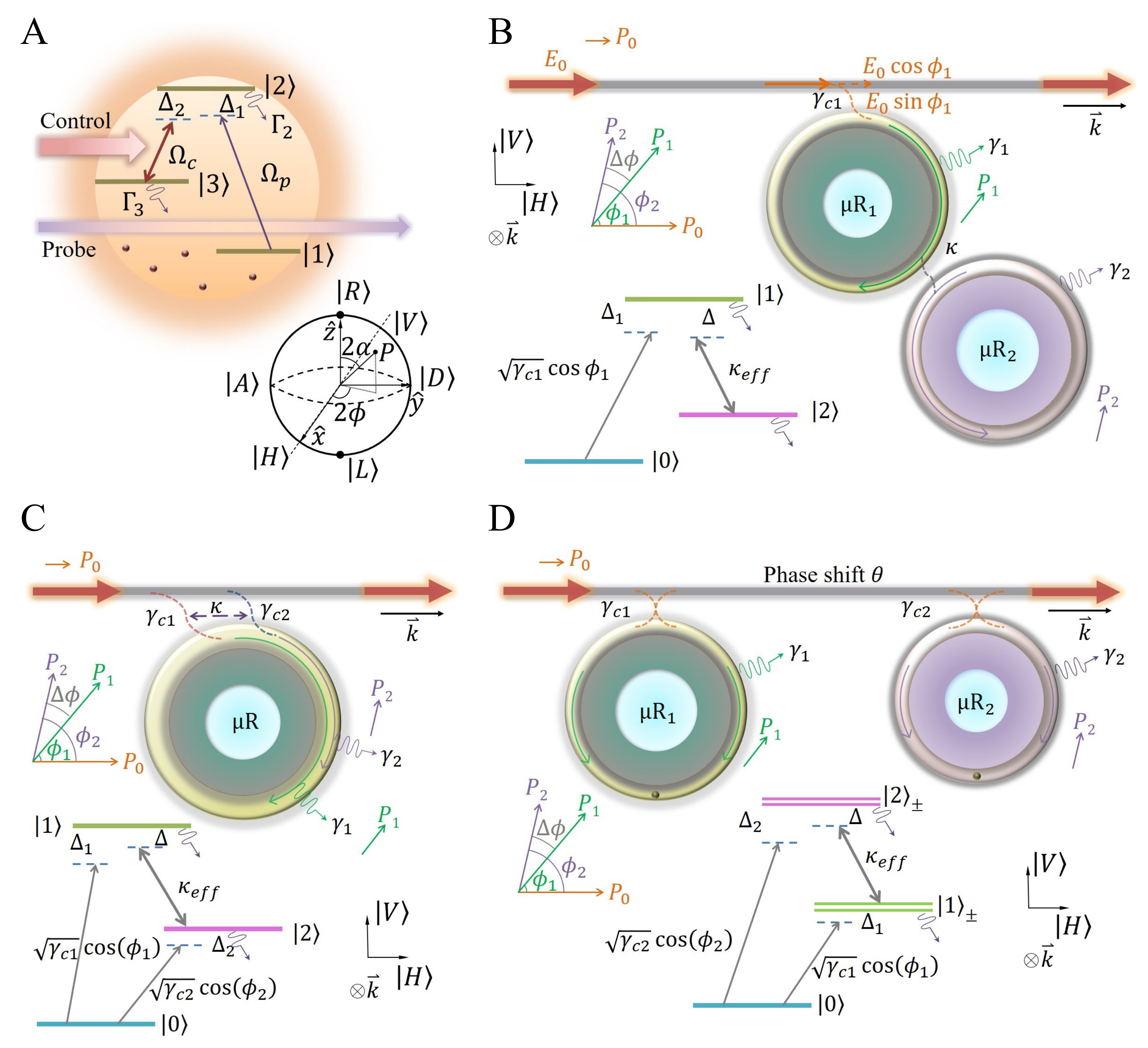}
\caption{Polarization
effects in different platforms for studying electromagnetically induced
transparency (EIT). (A) Atomic gas. Ground
state: $|1\rangle$, excited state: $|2\rangle$, metastable state: $|3\rangle$.
The control and probe light beams have Rabi frequencies $\Omega_{c}$ and
$\Omega_{p}$ respectively. The detuning between $|1\rangle\rightarrow|2\rangle$ ($|2\rangle\rightarrow|3\rangle$) and the probe (control) light is $\Delta_{1}$ ($\Delta_{2}$). For either probe or control light, an arbitrary polarization state $P$ as a superposition of the right and left circular polarization states ($|R\rangle$ and $|L\rangle$) can be represented on a Poincare sphere \cite{jones2016poincare}. The right circular, left circular, diagonal linear and anti-diagonal linear polarization states are related to the horizontal and vertical polarization states by $|L,R\rangle=\left(|H\rangle \pm i|V\rangle\right)/\sqrt{2}$, $|D,A\rangle=\left(|H\rangle \pm |V\rangle\right)/\sqrt{2}$. The components of the probe and control light that have a matching polarization will interact with the atomic systems and
induce EIT, while the mismatching components will be transparent to the system. (B) Directly coupled microresonators. $|0\rangle$, $|1\rangle$ and $|2\rangle$ represent the vacuum state, photons in $\mu$R$_{1}$ and photons in $\mu$R$_{2}$, respectively. The polarization orientations of quasi-TE or quasi-TM modes \cite{min2007perturbative} are shown in the inset, where $\vec{k}$ is the wavevector. With polarization mismatch, the effective coupling strength between the two resonator modes becomes $\kappa_{eff}=\kappa$cos$\left(\Delta\phi\right)$. (C) A single microresonator with two coupled modes. $|0\rangle$, $|1\rangle$ and $|2\rangle$ representations are similar to B.The two modes have disparate quality factors and different polarization states, with the effective coupling strength $\kappa_{eff}=\kappa$cos$\left(\Delta\phi\right)+\sqrt{\gamma_{c1}\gamma_{c2}}$cos$\left(\phi_{1}\right)$cos$\left(\phi_{2}\right)$. (D) Indirectly coupled microresonators with backscattering. With polarization mismatch, the effective coupling strength between $|1\rangle$ and $|2\rangle$ is given by $\kappa_{eff}=\left(\gamma_{c1}\gamma_{c2}\kappa_{a21}\kappa_{b12}e^{2i\theta}\right)^{1/4}[$cos$\left(\phi_{1}\right)$cos$\left(\phi_{2}\right)]^{1/2}$.}
\label{Fig1}
\end{figure*}

EIT originates from atomic/molecular systems, such as atomic gases (Fig.~\ref{Fig1}A), which are modeled as $\Lambda$-shape energy levels and are composed of a
ground state $|1\rangle$, an excited state $|2\rangle$ and a metastable state
$|3\rangle$. The decay rate of state $|3\rangle$ is much smaller than that of
state $|2\rangle$. The probe (pump) light beam induces the dipole transition
$|1\rangle\rightarrow|2\rangle$ ($|2\rangle\rightarrow|3\rangle$), while the
dipole transition $|1\rangle\rightarrow|3\rangle$ is forbidden. To generate
each dipole transition, certain linearly or circularly polarized light is
needed, whose polarization state $P$ is a superposition of the right and left polarization states ($|R\rangle$ and $|L\rangle$), i.e., $P=~$cos$\left(\alpha\right)|R\rangle+e^{-i2\phi}$sin$\left(\alpha\right)|L\rangle$, where $\alpha\in\left[0,\pi/2\right]$ and $\phi\in\left[0,\pi\right]$. Thus $P$ can also be represented by a Bloch sphere \cite{jones2016poincare} as shown in the inset of Fig.~\ref{Fig1}A. If the polarization state of the input light does not match the dipole
transition, then only the component with aligned polarization orientation will
interact with the atomic levels, while the rest will be noninteracting and
transparent to the system. As a result, the polarization of the pump light
will affect how much control light is effectively coupled to $|2\rangle
\rightarrow|3\rangle$, and thus determine the effective Rabi frequency
($\Omega_{c}$). As for the probe light, only the components with the matched
polarization will get involved in the EIT process, whereas the other component
will be transparent regardless of the coupling between levels $|2\rangle$ and
$|3\rangle$, and thus will raise the baseline over the whole transmission spectrum.

In a pair of directly coupled resonators (Fig.~\ref{Fig1}B), the level diagram takes on
a very similar form to that of the atomic system mentioned above, if we make the correspondence: $\Omega_{c}\leftrightarrow
\kappa$, $\Gamma_{2}\leftrightarrow\gamma_{1}+\gamma_{c1}$, $\Gamma
_{3}\leftrightarrow\gamma_{2}+\gamma_{c2}$, where $\kappa$ is the coupling strength between the two resonators, and $\gamma_{1,2}$ ($\gamma_{c1,c2}$) are the intrinsic (coupling) loss rates of the resonators $\mu$R$_{1}$ and $\mu$R$_{2}$ respectively. The ground state is now replaced
by the vaccum state, while the numbers of photons in
%TCIMACRO{\U{3bc}}%
%BeginExpansion
$\mu$%
%EndExpansion
R$_{1}$ and
%TCIMACRO{\U{3bc}}%
%BeginExpansion
$\mu$%
%EndExpansion
R$_{2}$ play the roles of the occupancy of levels $|2\rangle$ and $|3\rangle$, respectively.
Whispering gallery modes (WGMs) supported by resonators usually have quasi-TE or quasi-TM polarization states \cite{min2007perturbative}. To simplify the analysis, we consider that the input light also has a linear polarization state, and we denote the angle between $P_1$ ($P_2$) and $P_0$ as $\phi_1$ ($\phi_2$). When the waveguide mode is coupled to
%TCIMACRO{\U{3bc}}%
%BeginExpansion
$\mu$%
%EndExpansion
R$_{1}$, only the component $E_{0}$cos$\left(  \phi_{1}\right)  $ in the
orientation of $P_{1}$ will be coupled to the cavity mode, while the
perpendicular component $E_{0}$sin$\left(  \phi_{1}\right)  $ will be
transparent and elevate the transmission baseline. On the other hand, when the light couples from $\mu$R$_{1}$ to $\mu$R$_{2}$, only the component in the orientation of $P_{2}$ will be able to contribute to the mode in $\mu$R$_{2}$, while the component perpendicular to $P_{2}$ will not. The same process happens when the light couples from $\mu$R$_{2}$ to $\mu$R$_{1}$. Therefore, the polarization mismatch leads to a reduced coupling efficiency ($\kappa_{eff}=\kappa$cos$(\Delta\phi)$). Consequently, the figure of merit of EIT is reduced and the baseline in the transmission spectrum is raised. 

In the single resonator case (Fig.~\ref{Fig1}C), a high-$Q$ mode and low-$Q$ mode
overlapping in the frequency spectrum can be coupled to each other directly by mode profile overlap as well as indirectly via a waveguide. The
level diagram reveals that both modes are excited by the probe light so that the system is deviated from a perfect EIT model due to the additional absorption into the high-$Q$ mode. Considering different quasi-linear polarization states in the waveguide and the two modes, the effective
coupling strength is modified as $\kappa$cos$\left(\Delta\phi\right)+\sqrt{\gamma_{c1}\gamma_{c2}}$cos$\left(  \phi_{1}\right)$cos$\left(  \phi_{2}\right)$, where $\gamma_{c1}$ ($\gamma_{c2}$) denotes the coupling strength between the waveguide and mode 1 (mode 2), $\kappa$ represents the direct coupling strength between mode 1 and mode 2, $\phi_{1}$ ($\phi_{2}$) is the angle between the polarization of mode 1 (mode 2) and that of the input field, and $\Delta\phi=\phi_2-\phi_1$. Moreover, it has been
reported that the coresonant modes with different polarization can induce
transparency even without mode coupling \cite{bui2016experimental}.

The optical analogue of EIT can also be realized in indirectly coupled
resonators, where the phenomena of EIT and absorption can be controlled by the
chiral state of one of the resonators \cite{EITEP2019Wang}. In an indirectly coupled resonator
system (Fig.~\ref{Fig1}D), we consider
%TCIMACRO{\U{3bc}}%
%BeginExpansion
$\mu$%
%EndExpansion
R$_{1}$ and
%TCIMACRO{\U{3bc}}%
%BeginExpansion
$\mu$%
%EndExpansion
R$_{2}$ to be a high-$Q$ and a low-$Q$ resonator, respectively, both of which support
WGMs with backscattering. Each level of the cavity resonance is split into two
levels \cite{zhu2010chip,ozdemir2014highly} and can be tuned to be degenerate at the EPs \cite{peng2016chiral}. The effective coupling between the
modes in two cavities is given by $(\gamma_{c1}\gamma_{c2}\kappa_{a21}\kappa_{b12}e^{2i\theta
})^{1/4}[$cos$(\phi_1)$cos$(\phi_2)]^{1/2}$, which vanishes at one type of EP ($\kappa_{a21}=0$) and exists at
the other ($\kappa_{a21}\neq0$). The transition $|0\rangle\rightarrow
|1\rangle$ can be neglected when it is much smaller than the transition
$|0\rangle\rightarrow|2\rangle\rightarrow|1\rangle$. However, if $P_{1}$ is
different from $P_{0}$ and $\gamma_{c1}\gtrsim\gamma_{1}$, the polarization
of light passing
%TCIMACRO{\U{3bc}}%
%BeginExpansion
$\mu$%
%EndExpansion
R$_{1}$ can be greatly rotated, which significantly affects the transition
$|0\rangle\rightarrow|2\rangle$ and gives rise to a reduced absorption at
%TCIMACRO{\U{3bc}}%
%BeginExpansion
$\mu$%
%EndExpansion
R$_{2}$. Such polarization effect will not only reduce the efficiency of the
EIT configuration, but also lead to another kind of induced transparency phenomenon,
which we will discuss in detail. 

In all the above cases, the polarization
mismatch in the control light or mode coupling reduces the efficiency of EIT,
but does not break the fundamental conditions of EIT. Similarly, the EIT
efficiency is reduced by the polarization mismatch of the probe light in the first
two cases. Nevertheless, in the last two cases, the polarization mismatch
between the input light and the mode will induce fundamentally different
phenomena. 

\section*{Polarization induced transparency (PIT)}
The most intriguing polarization induced phenomenon can be seen from the indirect coupling scheme. Consider two indirectly coupled resonators ($\mu$R$_{1}$ and $\mu$R$_{2}$) supporting clockwise (CW) and counterclockwise (CCW) whispering gallery modes (WGMs) with quality factors of $Q_{1}$ and $Q_{2}$ ($Q_{1}\gg Q_{2}$). We follow the notation used in Fig.~\ref{Fig1}D. The quasi-linear polarization states
$P_{1}$ and $P_{2}$ of the CW modes \cite{min2007perturbative} in $\mu$R$_{1}$ and $\mu$R$_{2}$ form angles of $\phi_{1}$
and $\phi_{2}$ relative to the polarization orientation of the input light ($P_{0}$). In describing the full scattering properties of the system, we introduce the relationship between the input and output fields as%
\begin{equation}
	\begin{pmatrix}
		\lambda_x^{\prime} \\
		\lambda_y^{\prime} \\
		\rho_x^{\prime} \\
		\rho_y^{\prime}
	\end{pmatrix} =S\begin{pmatrix}
		\lambda_x \\
		\lambda_y \\
		\rho_x \\
		\rho_y
	\end{pmatrix}, \label{input_output}
\end{equation}
with $\lambda_{x(y)}$ and $\rho_{x(y)}$ being the $x (y)$ polarization components of the left- and right-incident field amplitudes, respectively. The $\lambda_{x(y)}^{\prime}$ and $\rho_{x(y)}^{\prime}$ are the $x(y)$ polarization components of the outcoming field amplitudes from the left and right ports, respectively. We now consider $t_{1,2}$ to be the transmission matrices of each individual resonator. The reflections to the left (right) are represented by $r_{1L}$ ($r_{1R}$) and $r_{2L}$ ($r_{2R}$) for $\mu$R$_1$ and $\mu$R$_2$, respectively. The scattering matrix can be written as
\begin{equation}
	\label{general_s}
	S = \begin{pmatrix}
		r_L & t_L\\
		t_R & r_R
		\end{pmatrix}, 
\end{equation}
with
\begin{align}
	r_L=&U_1^Tt_1^TU_2^T\left(1-r_{2L}U_2r_{1R}U_2^T\right)^{-1}r_{2L}U_2t_1U_1\nonumber\\
	&+U_1^{T}r_{1L}U_1,
\end{align}
\begin{equation}
	r_R=t_2U_2r_{1R}U_2^T\left(1-r_{2L}U_2r_{1R}U_2^T\right)^{-1}t_2^T+r_{2R},
\end{equation}
\begin{equation}
	t_R^T=t_L=U_1^Tt_1^TU_2^T\left(1-r_{2L}U_2r_{1R}U_2^T\right)^{-1}t_2^T,
\end{equation}
where $U_{1,2}$ are unitary matrices encoding the polarization mixing during the light propagation from the input port to $\mu {\rm R}_1$ and the propagation between $\mu {\rm R}_1$ and $\mu {\rm R}_2$ (due to, for example, polarization controllers inserted onto the waveguide).

To see the phenomena purely induced by polarization effects, we investigate a simple case that the resonators have
no backscattering on their surfaces and support degenerate WMGs, with the $S$ matrix
\begin{equation}
	\label{scattering}
	S = \begin{pmatrix}
		0 & (t_2 U_2 t_1 U_1)^T\\
		t_2 U_2 t_1 U_1 & 0
		\end{pmatrix}.
\end{equation}
$t_{1,2}$ can be calculated using the temporal coupled mode theory (TCMT)
\begin{equation}
	t_{1,2}=1-2i W_{1,2}^\dagger (\omega-H_{{\rm eff}\,1,2})^{-1} W_{1,2},
\end{equation}
where the effective Hamiltonian is 
\begin{equation}
	H_{{\rm eff}\,1,2}=\omega_{1,2}-i\gamma_{1,2}/2 - i W_{1,2} W_{1,2}^\dagger,
\end{equation}
and the coupling matrix is
\begin{equation}
	W_{1,2} = \sqrt{\frac{\gamma_{c1,2}}{2}}\begin{pmatrix}
		e^{i\delta_{1,2}} \cos \phi_{1,2} &
		e^{i\chi_{1,2}} \sin \phi_{1,2}\\
	\end{pmatrix}, \label{coupling}
\end{equation}
with $\delta_{1,2}$ and $\chi_{1,2}$ being the phases related to the coupling coefficients. The forward and backward transmission spectra can be obtained by solving Eqs.~(\ref{scattering})-(\ref{coupling}).

\begin{figure}%[ptb]
\includegraphics[width=1\linewidth]{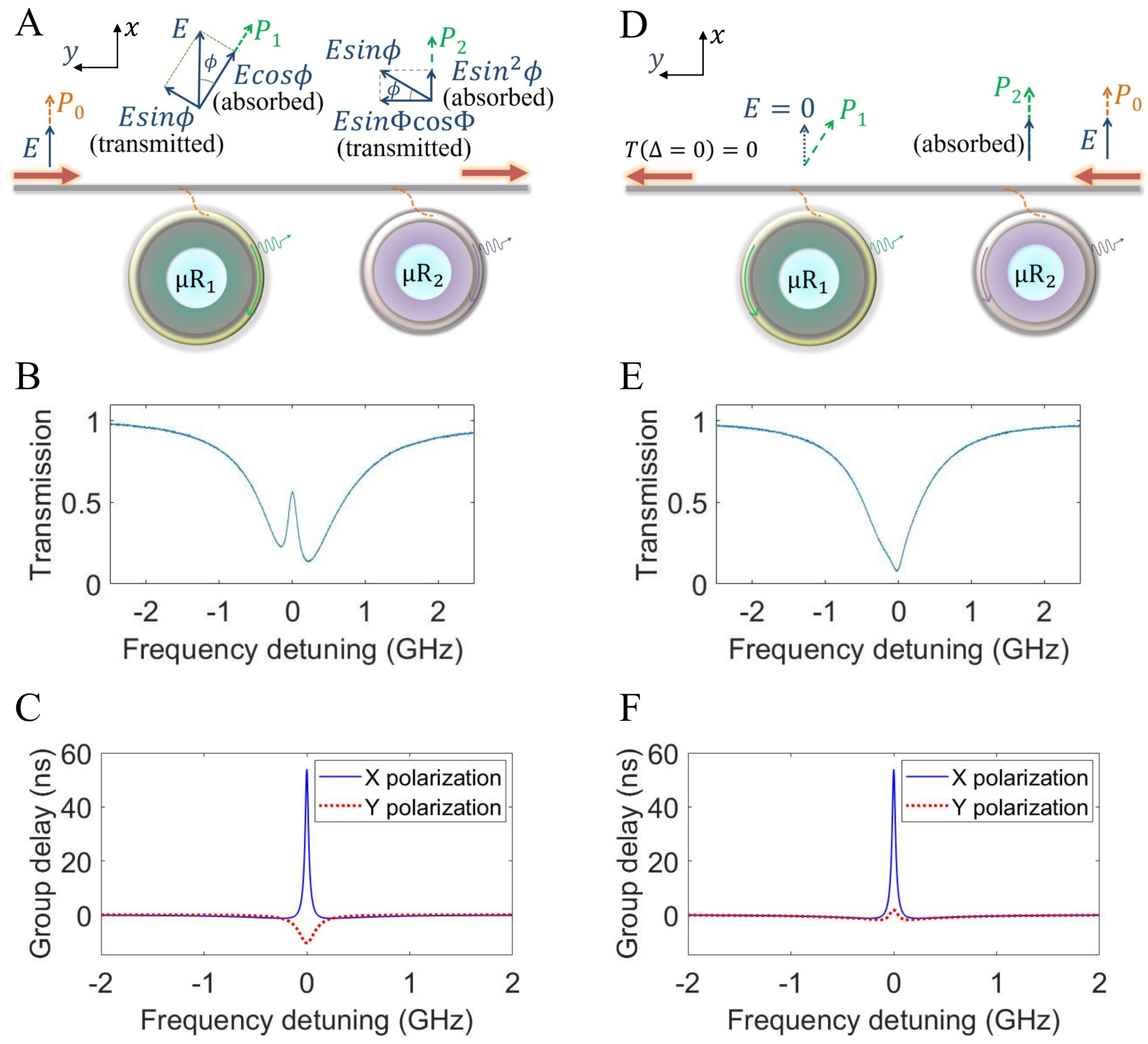}
\caption{Unidirectional polarization induced transparency (PIT). (A and D) Schematic diagrams of a
single mode waveguide coupled to two microresonators with no backscattering ($\mu$R$_{1}$ and $\mu$R$_{2}$). The vectors marked above the optical paths explain the polarization decomposition during light propagation in the case that both resonators are critically coupled to the waveguide. (B and E) Experimental results of (B) forward and (E) backward transmission spectra. (C and F) Calculated group delay for (C) forward and (F) backward propagation.}%
\label{Fig2}%
\end{figure}

To show polarization-induced phenomena, we design an experimental setup where a high-$Q$ microtoroid resonator ($\mu$R$_{1}$) and a low-$Q$ microtoroid resonator ($\mu$R$_{2}$) are coupled to a taper fiber waveguide. We investigate the case that $P_{0}$ aligns with $P_{2}$, which are both in $x$ direction. The angle between $P_{1}$ and $P_{2}$ ($\phi$) is set to be 0.25$\pi$ achieved by a polarization controller (PC) applied onto the waveguide between them. When only $\mu$R$_{2}$ is critically coupled to the taper, a
single Lorentzian dip appears in the transmission spectrum.
However, when $\mu$R$_{1}$ is also coupled to the taper (Fig.~\ref{Fig2}A), a narrow
transparency window appears in the forward transmission spectrum (Fig.~\ref{Fig2}B). This phenomenon originates from the polarization decomposition when the field travels from the waveguide ($E$) to $\mu$R$_{1}$, or vice versa. When $E$ encounters $\mu$R$_{1}$, only the component with the polarization orientation in the direction of $P_{1}$ interacts with the resonator, and passes with ratio $t_1$, while the perpendicular component gets fully transmitted. Thus the light passing $\mu$R$_{1}$ will have a polarization state significantly changed from $P_{0}$, which cannot be completely absorbed by $\mu$R$_{2}$. The modified absorption spectrum of the system is accompanied with a change of dispersion, based on the connection of real and imaginary parts of the response function governed by the Kramers-Kronig relations. The group delay of both $x$ and $y$ polarization components of the forwardly propagating field can be calculated by \cite{lu2018optomechanically}
\begin {equation}
    \tau_{x,y}=-\frac{d\left[arg\left(t_{x,y}\right)\right]}{d\omega},
\end {equation}
where $\omega$ is the frequency of the input optical field, and the transmission rates are related to the $S$ matrix in \eqref{scattering} by $t_x=S_{3,1}$ and $t_y=S_{4,1}$. The numerical results show that the $x$ polarization component of the output exhibits a large group delay within a narrow spectrum window, while the $y$ polarization component of the output shows group advance (Fig.~\ref{Fig2}C). Therefore, slow and fast light features are associated with different polarization states of the output light. The principle of the induced transparency phenomenon is different from EIT and EPAT, and thus we name it polarization induced transparency (PIT).

Furthermore, the PIT is unidirectional. In particular, when $\mu$R$_{2}$ is critically coupled to the taper, i.e., $\gamma_{c2}=\gamma_2$, the backward transmission spectrum displays a pure absorption dip (Fig.~\ref{Fig2} D and E), due to the fact that the field at zero detuning is fully absorbed by $\mu$R$_{2}$ before probing $\mu$R$_{1}$. In addition, the modulation on the group velocity is also found to be unidirectional, as the $y$ polarization component of the backwardly propagating light (when the polarization of the input is still in $x$ direction) exhibits group delay instead of group advance (Fig.~\ref{Fig2}F). The unidirectionality in transmission spectrum and dispersion uniquely associated with PIT can serve as a criterion for distinguishing between EIT and PIT in this scheme. It is worth noting that the unidirectionality of PIT does not violate reciprocity; the $S$ matrix here has reciprocity symmetry. This symmetry does not imply that, for a given input polarization, the total left to right output equals the total right to left output summed over polarizations.

\begin{figure}[ptb]
\includegraphics[width=1\linewidth]{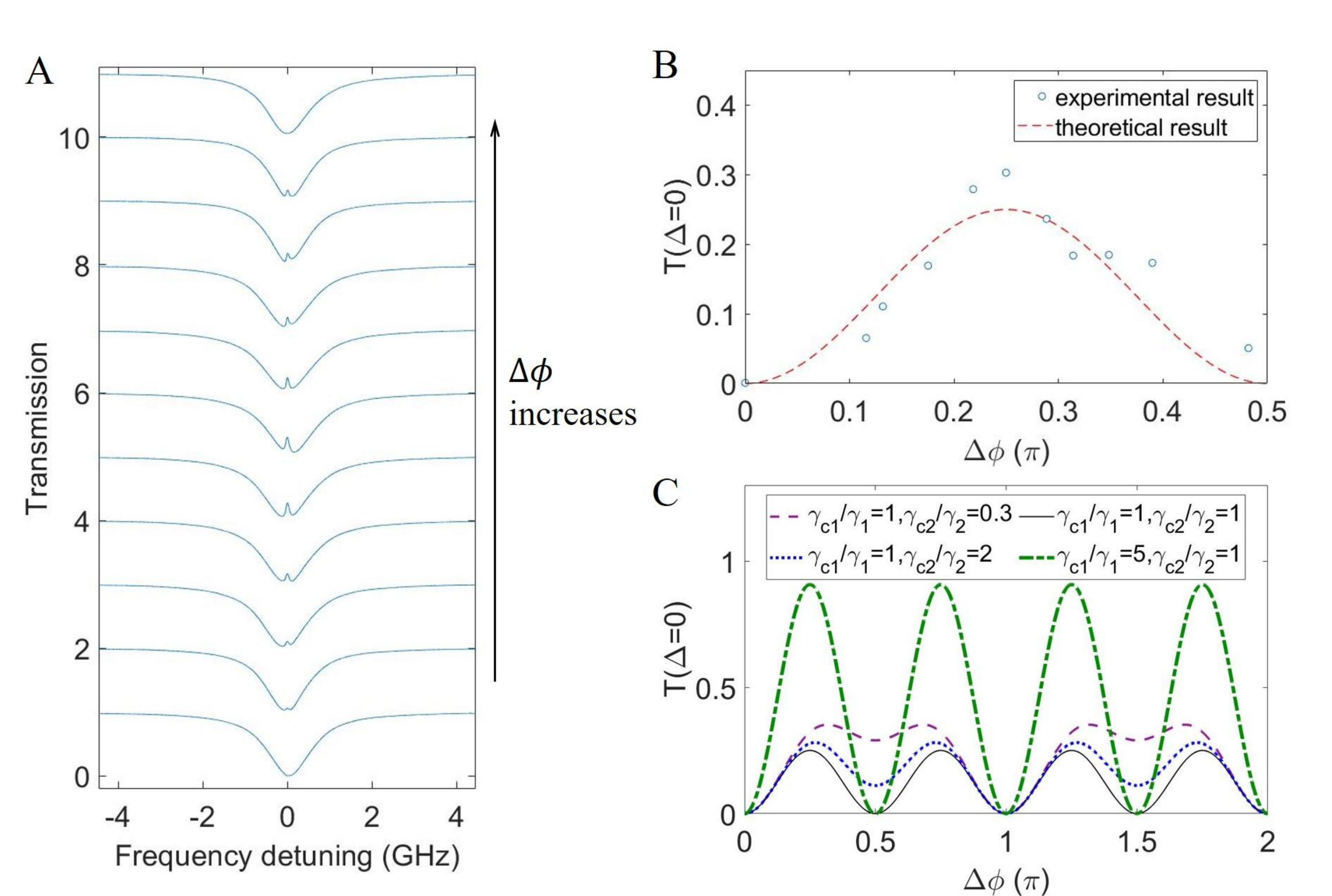}
\caption{Effects
of the polarization mismatch between the two resonators and the resonator-taper coupling
strengths on PIT. (A)
Experimentally measured transmission spectra of two indirectly coupled microtoroid resonators
($\mu$R$_{1}$: high-$Q$, $\mu$R$_{2}$: low-$Q$) as a function of the change of polarization of $\mu$R$_{1}$, which is kept at linear polarization and rotates by $\Delta\phi$
with respect to the polarization state of $\mu$R$_{2}$. The polarization state of $\mu$R$_{2}$ is aligned with that of the incident light. From bottom to top, $\Delta\phi$ increases from 0 to $\pi/2$. (B) Transmission at zero detuning versus the angle change of the polarization orientation of $\mu$R$_{1}$. The blue
circles are experimental result from A. The red dotted line is the
theoretical result with $\gamma_{c1}=\gamma_{1}$ and
$\gamma_{c2}=\gamma_{2}$. (C) Theoretical results of the transmission at zero detuning versus the change of polarization state of
$\mu$R$_{1}$ at different resonator-waveguide coupling strengths.}%
\label{Fig3}%
\end{figure}

We then study how the polarization state of $\mu$R$_{1}$ affects the forward transmission spectrum. The polarization of the input laser is controlled by a polarization controller (PC1). We apply another polarization controller (PC2) to the intermediate fiber between the two resonators, so that the polarization state of the light flowing through this region can be controlled manually. After initially aligning $P_{0}$ to $P_{2}$ by PC2, we can apply a linear rotation of PC1 and PC2 in opposite directions to equivalently adjust the $P_{1}$, while ensuring that $P_{0}$ and $P_{2}$ remain aligned (See Materials and Methods for details). $P_{1}$ is recorded by separately characterizing the transmission spectrum of $\mu$R$_{1}$. Starting from the condition $P_{1}=P_{2}=P_{0}$, we observe a narrow dip out of a wide absorption spectrum of $\mu$R$_{2}$. With the effective change of the orientation of $P_{1}$ (rotation angle $\Delta\phi$), the absorption rate of $\mu$R$_{1}$ is gradually reduced, accompanied by the appearance of narrow peaks in the spectrum (Fig.~\ref{Fig3}A). The peak at zero detuning undergoes oscillation with increase of $\Delta\phi$, reaching the maximum around $\Delta\phi=0.25\pi$ (Fig.~\ref{Fig3}B).

The effect of polarization can also be modulated by the waveguide-resonator coupling strengths $\gamma_{c1}$ and $\gamma_{c2}$. This is shown by studying the variation of transmission at the zero detuning
versus the change of $P_{1}$ under different $\gamma_{c1}$ and $\gamma_{c2}$. When the coupling strength between the high-$Q$ resonator $\mu$R$_{1}$ and the taper is increased and pushed into the overcoupled regime, one can find a higher transparency peak (green dashed curve in Fig.~\ref{Fig3}C) compared to the undercoupling regime (black solid curve in Fig.~\ref{Fig3}C). This owes to the fact that the transmission coefficient $t_1(\Delta)=\frac{i\Delta-\left(\gamma_1-\gamma_{c1}\right)/2}{i\Delta-\left(\gamma_1+\gamma_{c1}\right)/2}$ at zero detuning ($\Delta=0$) becomes negative in the strong coupling regime ($\gamma_{c1}>\gamma_1$), introducing a $\pi$ phase shift to the $P_1$ component of the transmitted field in the waveguide, which significantly rotates the polarization of the total field passing $\mu$R$_{1}$. On the other hand, PIT is also influenced by the coupling strength between $\mu$R$_2$ and the waveguide. Among all coupling conditions, the highest peak appears around
$\Delta\phi=\pi/4$, and a local minimum shows up at $\Delta\phi=\pi/2$, namely $P_{1}$ is
perpendicular to $P_{2}$ and $\mu$R$_{1}$ is decoupled from the optical path (Fig.~\ref{Fig3}C). Yet the contrasts of the transparency
window, which mark the efficiency of PIT, are smaller in the cases of undercoupling and overcoupling than in the critical coupling case. Thus based on the discussion above, PIT is optimized when $\mu$R$_{1}$ is overcoupled to the taper and $\mu$R$_{2}$ is critically coupled to the taper.

\section*{Hybrid system for EIT and PIT}

\begin{figure}%[ptb]
\centering
\includegraphics[width=1\linewidth]{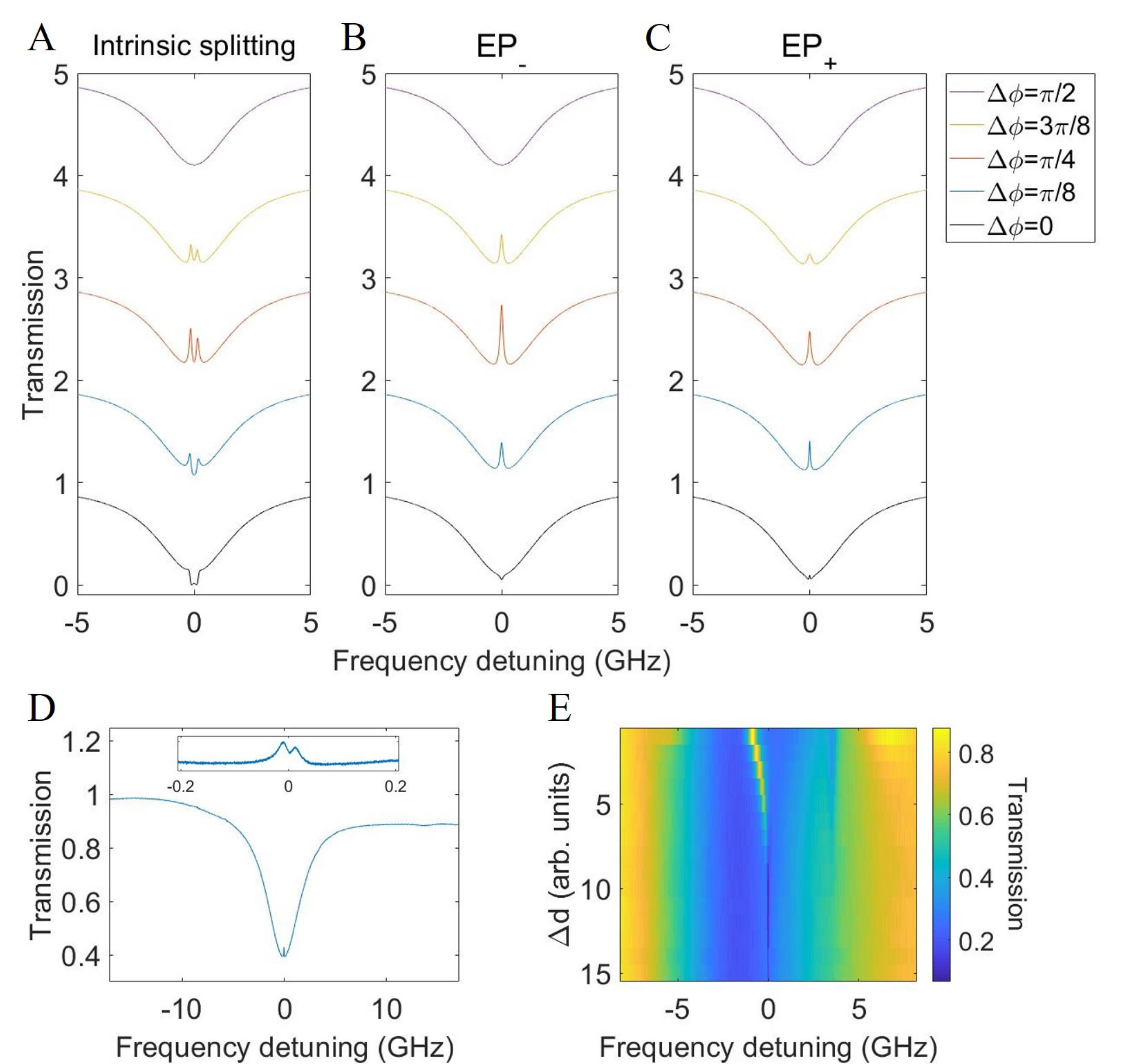}
\caption{Polarization effect in indirectly coupled resonators with backscattering.
(A-C) Theoretical results in intrinsic splitting case (A), EP$_{-}$ case (B) and EP$_{+}$ case (C), with different relative polarization angles ($\Delta\phi$). The transmission
values are shifted by 1 in each curve for visual comparison. (D) Experimentally obtained transmission spectrum when $\mu$R$_{1}$ has intrinsic splitting. The inset shows a close-up of the transmission spectrum around the zero detuning. (E) Experimentally obtained transmission spectra with the change of the gap between $\mu
$R$_{1}$ and the taper ($\Delta d$).}%
\label{Fig4}%
\end{figure}

We finally investigate indirectly coupled resonators with backscattering, where EIT and PIT appear simultaneously. By steering $\mu$R$_{1}$ to EPs, transparency or absorption occurs depending on the type of EPs classified by the chirality of eigenstates \cite{EITEP2019Wang}. For EP$_{-}$ where the eigenmode is in the CCW direction and has chirality -1, the interference is “swicthed off” resulting in exceptional-point-assisted absorption (EPAA). For EP$_{+}$ at which the eigenmode is in the CW direction with chirality +1, the destructive interference leads to EPAT. Here we find that the polarization mismatch $\Delta\phi$ could singificantly modify the transmission spectra. With the intrinsic splitting of $\mu$R$_{1}$, under weak coupling between $\mu$R$_{1}$ and the taper, the transmission shows a splitting absorption window when $\Delta\phi=0$, but exhibits a splitting transparency window when $\Delta\phi=\pi/4$ (Fig.~\ref{Fig4}A). When $\mu$R$_{1}$ is steered to EP$_{-}$ (or EP$_{+}$), the lineshape of EPAA (or EPAT) appears when $\Delta\phi=0$ (Fig.~\ref{Fig4} B and C). But with the polarization mismatch, a large transparency window can be induced in the forward transmission spectrum in both the cases of EPAA and EPAT. 

In experiments, we choose a microtoroid ($\mu$R$_{1}$) and a microdisk ($\mu$R$_{2}$)
resonator with strong backscattering and polarization mismatch. With intrinsic mode splitting in both resonators, a
transparency window with splitting is observed (Fig.~\ref{Fig4}D). The peak goes larger with increased coupling strength between $\mu$R$_{1}$ and the fiber taper (Fig.~\ref{Fig4}E). 

\section*{Discussions}

The physical phenomena and processes discussed above shed light on the distinction between EIT and PIT.
First, the all-optical analogue of EIT in linear optical systems is the direct result
of interference in the optical paths and has the $\Lambda$-type level structure, whereas the occurrence of PIT has no relevance to interference effects. Second, while EIT depends on large intermodal coupling, PIT occurs in the absence of it and can display a large transparency window based on the strong polarization rotation effect enabled by the microresonators. Third, PIT is accompanied by the unidirectional behavior, while EIT occurs for transmission in both directions.

Such a clarification is important not only in terms of accuracy of physics concepts, but also from the perspective of applications. Slow light application relies on group delay in optical signal, which can be realized by the all-optical analogue of EIT, EPAT, optomechanically induced transparency (OMIT) \cite{weis2010optomechanically,safavi2011electromagnetically,dong2012optomechanical,lu2018optomechanically} and Brillouin-scattering-induced transparency (BSIT) \cite{kim2015non,dong2015brillouin}, etc. With different mechanism from EIT, PIT offers an alternative approach to manipulate the group index of optical media for control of slow light which is direction- and polarization-dependent. Furthermore, the unidirectionality associated with PIT enables directional control of light transport without the need of any nonlinear elements or external control, which can potentially benefit optical information processing in on-chip all-optical devices, systems and networks.

\section*{Materials and Methods: Control of the polarization of one resonator by a polarization controller}

In experiment for Fig.~\ref{Fig3} in the main text, we intend to rotate the polarization of the mode in $\mu$R$_{1}$ without physically changing the optical structure. Therefore, after initially aligning $P_0$ to $P_2$ by PC2, we apply a linear rotation of PC1 and PC2 in opposite directions. We now prove that this method can equivalently adjust $P_1$ without breaking the alignment between $P_0$ and $P_2$.
The rotation of $P_1$ and $P_2$ in opposite directions but by same degree ensures that any rotation of PC2 described by a rotation matrix $U$ is accompanied with a rotation $U^{\dagger}$ on PC1, so that the output vector from the right port becomes 
\begin{equation}
	\begin{pmatrix}
		\rho_x^{\prime} \\
		\rho_y^{\prime}
	\end{pmatrix} =t_{2}Ut_{1}U^{\dagger}\begin{pmatrix}
		\lambda_x \\
		\lambda_y \\
	\end{pmatrix}.
\end{equation}
This operation can be regarded as applying a rotation to $t_{1}$, that is
\begin{equation}
	Ut_{1}U^{\dagger}=1-2 i (UW_{1}^\dagger) (\omega-H_{{\rm eff}\,1})^{-1} (W_{1}U^{\dagger}),
\end{equation}
which is equivalent to the rotation of the coupling matrix $W$, or the polarization state of $\mu$R$_{1}$, by $U$. Thus an arbitrary polarization mismatch between the two resonators can be chosen.

\section*{Acknowledgments}
This work was supported by the NSF grant No. EFMA1641109, ARO grant No. W911NF1710189, and DARPA under grant HR00111820042. A.D.S. acknowledges the support of NSF grant No. DMR-1743235. L.J. acknowledges the support of the Packard Foundation (2013-39273). C.W. acknowledges the fellowship support through the McDonnell International Scholars Academy.

%\begin{thebibliography}{9}                                                                                                %
%\section*{Reference}
%\bibliographystyle{natbib}
\nocite{*}
%\bibliography{MyCollection.bib}

\end{document}